\documentclass[a4paper,12pt]{article}
\usepackage[cp1251]{inputenc}
\usepackage{amsfonts,longtable}
\usepackage{amsmath}
\usepackage{amssymb}
\usepackage{textcomp}
\usepackage{graphicx}
\usepackage[pdftex]{color}
\usepackage[square,numbers,comma]{natbib}
\usepackage[pdftex]{hyperref}
\hypersetup{
             final,
             colorlinks=true,
             linkcolor=blue,
             citecolor=red
}

\begin{document}
\begin{center}
{\Large\bf QUANTUM WELL BASED ON GRAPHENE AND\\
\vspace{0.1cm} NARROW-GAP SEMICONDUCTORS}

\vspace{0.5cm}

P. V. Ratnikov\footnote{\url{ratnikov@lpi.ru}} and A. P. Silin

\textit{Tamm Theoretical Department of the Lebedev Physics
Institute, RAS\\ Leninskii pr. 53, 119991, Moscow, Russia}
\end{center}

\vspace{0.2cm}

\begin{list}{}
{\rightmargin=0cm\leftmargin=2cm}\item\fontsize{11pt}{10pt}\selectfont{
{\bf Abstract.} We consider the energy spectrum of the planar
quantum well which consisted of two ribbons of narrow-gap
semiconductors and a graphene ribbon between ones. It is shown that
the gapless mode appears only in case of inverted narrow-gap
semiconductors. Spin splitting of the energy spectrum for a
nonsymmetric quantum well is calculated taking into account a
specificity of graphene. We investigate interface states and optical
transitions. It is shown that the optical transitions are possible
only with a conservation of a parity.}

\vspace{0.5cm}

\end{list}

\vspace{0.5cm}

Submitted to: Bulletin of the Lebedev Physics Institute

\vspace{0.5cm}

After the first experimental investigation of graphene, a monatomic
layer of carbon atoms forming a regular hexagonal lattice
\cite{Novoselov, Zhang}, an intensive activity has arisen in
different directions. In particular, a multitude of works on
electronic properties of narrow graphene ribbons with nanometer
sizes (nanoribbons) of late three years were made. Among early ones,
there were works on study of electron states of the graphene ribbons
using the Dirac equation with the appropriate boundary conditions
\cite{Brey1, Brey2}. The electronic properties of the graphene
nanoribbon (GNR) depend strongly on size and geometry, as well as on
whether hydrogen atoms were deposited on free atomic orbitals of
carbon atoms at edges of the GNR \cite{Son}. In their transport
properties, the GNRs are very similar to carbon nanotubes, since a
free motion is one-dimensional (1D) in both cases \cite{Saito,
Ando}.

The boundary conditions for nanotubes, in which charge carriers are
described by the Dirac equation, were discussed in detail in Ref.
\cite{McCann}. Two types of the boundary conditions correspond to
two types of edges of the graphene ribbon, namely zigzag and
armchair \cite{Brey1}. A formation of the zero-energy surface states
is characteristic for the GNRs with the former type of edges. The
energy spectrum is gapless (metallic) for the latter type of edges,
when the ribbon width $d=(3M+1)a$, where $M$ is an integer and
$a=\sqrt{3}a_0$ is the lattice constant ($a_0$ is the interatomic
distance), and insulating otherwise \cite{Brey2}.

Recently, a transistor was made on the basis the GNR with the 2-nm
width and 236-nm length (nanoribbons with the 10--60-nm width were
also investigated) \cite{Wang}. The GNR was taken so narrow to
guarantee a wide semiconducting energy gap for room  temperature
operation of the transistor. However, it loses slightly a
compactness to the transistor based on a graphene quantum dot with
the 30-nm  diameter \cite{Ponomarenko}.

\newpage

A planar contact of the GNR with zigzag edges with metal ribbons has
been investigated in Ref. \cite{Blanter} within the tight-binding
approximation. In our paper, we consider the size quantization of
massless charge carriers in the GNR having armchair edges with the
width $d\neq(3M+1)a$ between narrow-gap semiconductor ribbons.
Changing narrow-gap semiconductors, we can change potential
barriers heights which are the band gaps of the narrow-gap
semiconductors, what opens additional possibilities for the energy
band-gap engineering \cite{Han}.

The potential barrier height for the single GNR has been calculated
in Ref. \cite{Okada} within the local spin density approximation. It
equals 2.33 and 3.33 eV for the armchair and zigzag edges,
respectively. These values is one order higher than a typical band
gap of the narrow-gap semiconductors.

We carried out calculations within the Dirac model \cite{Volkov1}.
Let us direct $X$ axis parallel to interfaces of the narrow-gap
semiconductors and graphene, $Y$ axis transversely to a
heterostructure plane, and $Z$ axis transversely to $X$ and
$Y$ axes (\hyperlink{fig1}{Fig. 1}). In a two-band approximation, a heterostructure composed
of narrow-gap semiconductors\footnote{We can use a finite-gap graphene modification as a narrow-gap
semiconductor \cite{Peres}.} 
and graphene is described by the
Dirac equation. Half the band gap $\Delta$, the work function $V$,
and the matrix element for the rate of interband transitions $u$ are
constant and change only on the heterostructure boundaries. For
chosen orientation of axes in given case, the Dirac equation has
following form
\begin{equation}\label{1}
\widehat{H}_D\Psi\equiv\left\{u_i\gamma^0\gamma^1\widehat{p}_x+u_i\gamma^0\gamma^3\widehat{p}_z
+\gamma^0\Delta_i+V_i\right\}\Psi=E\Psi,
\end{equation}
where $\gamma^0=\begin{pmatrix}I&0\\0&-I\end{pmatrix}$ and
$\boldsymbol{\gamma}=\begin{pmatrix}0&\boldsymbol{\sigma}\\-\boldsymbol{\sigma}&0\end{pmatrix}$
are the Dirac $\gamma$-matrices, $I$ is the $2\times2$ unit matrix,
$\boldsymbol{\sigma}$ are the Pauli matrices, $\widehat{\bf
p}=-i\boldsymbol{\nabla}$ is the momentum operator, and $\hbar=1$.

It is simply verified that the operator\footnote{One should
distinguish the operator $\widehat{P}$ from the helicity operator
$\widehat{h}$, introduced for \textit{unbounded} graphene
\cite{Neto}, which is the operator of a projection of pseudospin
onto a free motion momentum. The free motion is 1D in the planar
heterostructure and the 1D operator $\widehat{h}$ (written in the
$4\times4$ matrix form) does not commute with $\widehat{H}_D$.
Therefore, an eigenvalue of $\widehat{h}$ does not conserve. The
conserved quantum number is the eigenvalue of the operator
$\widehat{P}$.}
\begin{equation}\label{2}
\widehat{P}=i\gamma^0\gamma^3\gamma^1
\end{equation}
commutes with Hamiltonian $\widehat{H}_D$.

To clear up the sense of the operator $\widehat{P}$ we present it in
the form
\begin{equation}
\widehat{P}=i\gamma^0\widehat{\Lambda}_y,
\end{equation}
where $i\gamma^0$ is the inversion operator,
$\widehat{\Lambda}_y=e^{-i\frac{\pi}{2}\Sigma_y}=-i\Sigma_y$ is the
operator of the rotation about the $Y$ axis by angle $\pi$ (see,
e.g., \cite{Akhiezer}), and
$\Sigma_y=\begin{pmatrix}\sigma_y&0\\0&\sigma_y\end{pmatrix}$. From
this, one can see that the operator $\widehat{P}$ plays the role of
the pseudoparity operator introduced for the layered narrow-gap
semiconductor heterostructures \cite{Idlis}.

\newpage

\begin{center}
\hypertarget{fig1}{}
\includegraphics[width=16cm]{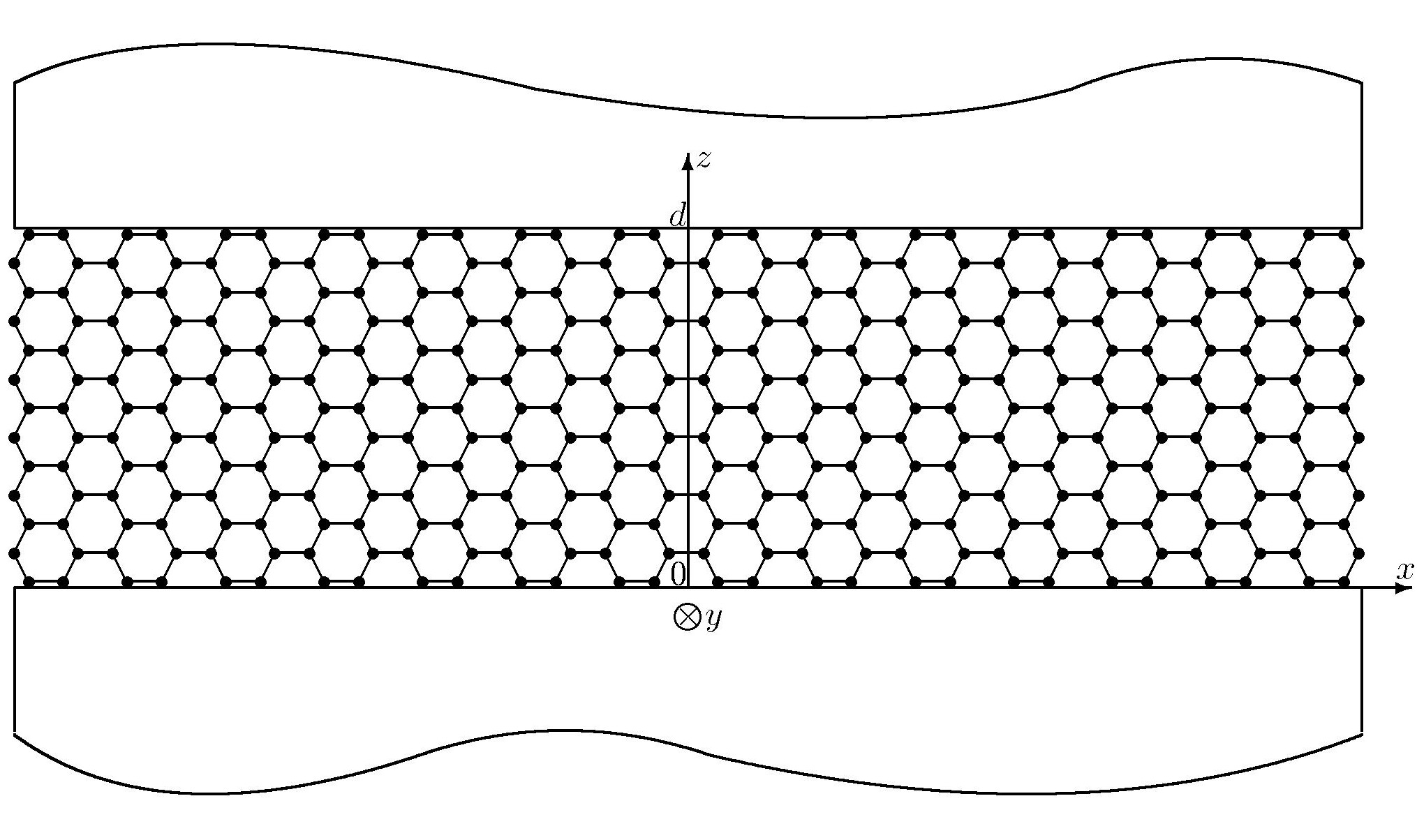}

Fig. 1. \textit{Considered heterostructure: the GNR between two the
narrow-gap semiconductor ribbons (here $Y$ axis is directed from us;
an atomic structure of narrow-gap semiconductors is not shown).}
\end{center}
\vspace{0.5cm}

The eigenvalues $\lambda=\pm1$ of the operator $\widehat{P}$
determine sign of a spin projection onto $Y$ axis. It can be
verified by the calculation of the mean value of spin as it was made
in Ref. \cite{Idlis}. Hereinafter, we will refer to $\lambda$ as the
parity. As it was emphasized in our previous paper \cite{Ratnikov1},
the charge carriers in graphene are the chirality massless Dirac
fermions (spin of particle is aligned with its momentum or is
opposed to it). As it will be shown below, there is no the gapless
mode in the quantum well considered. The charge carriers in the GNR
of a finite width bounded by layers of usual (noninverted)
narrow-gap semiconductors, have nonzero effective mass and its
energy spectrum is similar to the semiconductor spectrum with the
finite band gap.

We find a solution to equation \eqref{1} in a class of functions
being the eigenfunctions of the pseudoparity operator \eqref{2}.
Components of the eigen bispinor interconnect by relations
\begin{equation}
\begin{split}
&\psi_{\lambda2}=i\lambda\psi_{\lambda1},\\
&\psi_{\lambda4}=-i\lambda\psi_{\lambda3}.
\end{split}
\end{equation}
Now the Dirac equation \eqref{1} is written in the form
\begin{equation}
\left\{u_i\gamma^0\gamma^3\widehat{p}_z-i\lambda\gamma^3u_ik_x+\gamma^0\Delta_i+V_i\right\}\Psi_\lambda=
E_\lambda\Psi_\lambda.
\end{equation}
Two independent components of the eigen bispinor satisfy the system
of equations
\begin{equation}\label{6}
\begin{aligned}
\left(-iu_i\frac{d}{dz}+i\lambda
u_ik_x\right)\psi_{\lambda1}=\left(E_\lambda+\Delta_i-V_i\right)\psi_{\lambda3},\\
\left(-iu_i\frac{d}{dz}-i\lambda
u_ik_x\right)\psi_{\lambda3}=\left(E_\lambda-\Delta_i-V_i\right)\psi_{\lambda1}.\\
\end{aligned}
\end{equation}
We find the solution in three ranges in the form

1) $z<0$: \, $\psi_{\lambda1}=A_1e^{k_1z},\,
\psi_{\lambda3}=A_3e^{k_1z}$,

2) $0<z<d$: \,
$\psi_{\lambda1}=B_1e^{ikz}+\widetilde{B}_1e^{-ikz},\,
\psi_{\lambda3}=B_3e^{ikz}+\widetilde{B}_3e^{-ikz}$,

3) $z>d$: \, $\psi_{\lambda1}=C_1e^{-k_3z},\,
\psi_{\lambda3}=C_3e^{-k_3z}$,

\hspace{-0.65cm}respectively, for the narrow-gap semiconductor with
$\Delta_1,\, V_1,\, u_1$, graphene with $\Delta_2=0,\, V_2=0,\,
u_2$, and the narrow-gap semiconductor with $\Delta_3,\, V_3,\,
u_3$.

Using the boundary conditions of continuity of $\sqrt{u}\Psi$
\cite{Silin1}, one can exclude constants $A_1, A_3, C_1, C_3, B_3,
\widetilde{B}_3$ and obtain the system of equations on coefficients
$B_1$ and $\widetilde{B}_1$\footnote{The system of equations in
$B_3$ and $\widetilde{B}_3$ is analogous to \eqref{7}.}
\begin{equation}\label{7}
\begin{aligned}
&\left(\frac{u_2k+i\lambda u_2k_x}{E_\lambda}+\frac{iu_1k_1-i\lambda
u_1k_x}{E_\lambda+\Delta_1-V_1}\right)B_1+
\left(\frac{-u_2k+i\lambda u_2k_x}{E_\lambda}+\frac{iu_1k_1-i\lambda
u_1k_x}{E_\lambda+\Delta_1-V_1}\right)\widetilde{B}_1=0,\\
&\left(\frac{u_2k+i\lambda u_2k_x}{E_\lambda}-\frac{iu_3k_3+i\lambda
u_3k_x}{E_\lambda+\Delta_3-V_3}\right)B_1e^{ikd}+
\left(\frac{-u_2k+i\lambda u_2k_x}{E_\lambda}-\frac{iu_3k_3+i\lambda
u_3k_x}{E_\lambda+\Delta_3-V_3}\right)\widetilde{B}_1e^{-ikd}=0.\\
\end{aligned}
\end{equation}

The dependence of energy on the allowed wave number $k$ and the wave
number of the free motion $k_x$ is given by relation
\begin{equation}\label{8}
E_\lambda=\pm u_2\sqrt{k^2+k^2_x}.
\end{equation}
The gapless mode corresponds to $k=0$. Then the energy spectrum is
linear one $E_\lambda=\pm u_2|k_x|$. However, from \eqref{7}, one
can see that then $B_1+\widetilde{B}_1=0$ and
$B_3+\widetilde{B}_3=0$, what means that $\Psi_\lambda\equiv0$,
therefore this solution is not physical one.

The gapless mode in the quantum well based on graphene is possible
if graphene is bounded by the inverted narrow-gap semiconductors
\cite{Volkov2} in which $\Delta_1\Delta_3<0,\, u_1=u_2=u_3=u$ and
$V_1=V_2=V_3=0$. The wave function is constant in the range $0<z<d$
at $k=0$, namely, $\psi_{\lambda1}=D_1,\, \psi_{\lambda3}=D_3$.
After a substitution of this solution in \eqref{6}, we obtain
$D_3=\pm iD_1$ and $E_\lambda=\pm\lambda u|k_x|$. Joining solutions
on boundaries $z=0$ and $z=d$, we find the wave function

1) $\Psi^{(\pm)}_\lambda=C\widetilde{\Psi}^{(\pm)}_\lambda
e^{\mp\frac{\Delta_1}{u}z}$,

2) $\Psi^{(\pm)}_\lambda=C\widetilde{\Psi}^{(\pm)}_\lambda$,

3) $\Psi^{(\pm)}_\lambda=C\widetilde{\Psi}^{(\pm)}_\lambda
e^{\mp\frac{\Delta_3}{u}(z-d)}$,

\hspace{-0.65cm}where
\begin{equation*}
\widetilde{\Psi}^{(\pm)}_\lambda=\begin{pmatrix}1\\i\lambda\\
\pm i\\ \pm\lambda\end{pmatrix}.
\end{equation*}
A constant $C$ is determined from the normalizing condition
\begin{equation}\label{9}
\int\limits_{-\infty}^\infty dz \Psi^\dagger_\lambda\Psi_\lambda=1,
\end{equation}
\begin{equation*}
C=\frac{1}{2}\left[\frac{u}{2}\left(\frac{1}{|\Delta_1|}+\frac{1}{|\Delta_3|}\right)+d\right]^{-1/2}.
\end{equation*}
The case $D_3=iD_1$ corresponds to $\Delta_1<0,\, \Delta_3>0$, and
$E_\lambda=\lambda uk_x$. For $D_3=-iD_1$, we obtain $\Delta_1>0,\,
\Delta_3<0$, and $E_\lambda=-\lambda uk_x$.

We consider below the case of the noninverted narrow-gap
semiconductors with $\Delta_{1,3}>0$ when there no the gapless mode.
It is simply obtained the equation in $k$ corresponded to the
equation investigated in Refs. \cite{Andushin1, Andushin2}
\begin{equation}\label{10}
\tan
kd=\frac{A(C+\widetilde{C})}{1+B(C-\widetilde{C})-C\widetilde{C}},
\end{equation}
where
\begin{equation*}
A=\frac{u_2k}{E_\lambda},\, B=\frac{\lambda u_2k_x}{E_\lambda},\,
C=\frac{u_1k_1-\lambda u_1k_x}{E_\lambda+\Delta_1-V_1},\,
\widetilde{C}=\frac{u_3k_3+\lambda u_3k_x}{E_\lambda+\Delta_3-V_3},
\end{equation*}
\begin{equation*}
u^2_1k^2_1=\Delta^2_1-(E_\lambda-V_1)^2+u^2_1k^2_x,\,
u^2_2k^2=(E_\lambda-V_2)^2-\Delta^2_2-u^2_2k^2_x,\,
u^2_3k^2_3=\Delta^2_3-(E_\lambda-V_3)^2+u^2_3k^2_x.
\end{equation*}

In general case, equation \eqref{10} is sufficient awkward therefore
we restrict oneself to a consideration of the particular cases for
the defined types of the quantum wells.

{\bf\textit{The symmetric quantum well.}}

We refer to such quantum well in which $V_1=V_3=V,\,
\Delta_1=\Delta_3=\Delta,\, \Delta_2=0,\, V_2=0,\,
u_1=u_3=\overline{u},\, u_2=u$ as the symmetric quantum well.
Equation \eqref{10} is reduced to the form
\begin{equation}\label{11}
\tan kd=\frac{u\overline{u}k_1k}{E(E-V)-u\overline{u}k^2_x}.
\end{equation}
The dependence of energy on $\lambda$ vanishes, therefore, there is
no parity splitting. Equation \eqref{11} reduces to the most simple
form at $V=0$ and $u=\overline{u}$
\begin{equation}\label{12}
\tan kd=\frac{\sqrt{\Delta^2-u^2k^2}}{uk}.
\end{equation}
From the last equation, one can simply find the number $n_{e(h)}$ of
branches in the electron (hole) size-quantization spectrum (in given
case $n_e=n_h=n$). This number is to be to valid an
inequality\footnote{It is interesting to note that there exists the
solution to equation \eqref{12} in the case $\frac{d\Delta}{u}=\pi
n$. This corresponds to a level lying on boundary with the
continuous spectrum.}
\begin{equation}\label{13}
\pi(n-1)\leq\frac{d\Delta}{u}<\pi n.
\end{equation}

It is evident from the graphical solution of equation \eqref{12}
that the theorem of nonrelativistic quantum mechanics remains valid
which states that there exists at least one level in an arbitrarily
shallow 1D symmetric quantum well\footnote{In the nonsymmetric
quantum well as in a nonrelativistic case (see, e.g.,
\cite{Landau}), there can exist no one level.}.

{\bf\textit{The nonsymmetric quantum well of the first type.}}

So we refer to the quantum well with $\Delta_1=\Delta_3=\Delta,\,
\Delta_2=0,\, V_1=-V_3=V,\, V_2=0,\, u_1=u_3=\overline{u},\, u_2=u$.
Then, $\lambda$ explicitly enters into equation \eqref{10} as
$\lambda k_x$. Hence, parity splitting (spin splitting) arises.
Energy extrema are not at $k_x=0$ and lie at $k_x=\lambda k^*_x$.
Hence, an energy depends on $k_x-\lambda k^*_x$. We suppose that
$|V|\ll\Delta$. In the lowest order of $V$, the dependence of energy
on a momentum $k_x$ is given by expression \eqref{8}
\begin{equation}\label{14}
E^{(0)}_\lambda=\pm u\sqrt{k^{(0)2}+k^2_x},
\end{equation}
where $k^{(0)}$ is the root of equation \eqref{11} at $V=0$
\begin{equation}\label{15}
\tan
k^{(0)}d=\frac{k^{(0)}\sqrt{\Delta^2-u^2k^{(0)2}+(\overline{u}^2-u^2)k^2_x}}{uk^{(0)2}+(u-\overline{u})k^2_x}.
\end{equation}
In the next order of $V$, we have\footnote{The quadratic dependence
of an energy on a momentum, which is valid at large values of
$\Delta_2$, was used in Ref. \cite{Andushin1}.}
\begin{equation}\label{16}
E^{(1)}_\lambda=\pm u\sqrt{k^{(0)2}+(k_x-\lambda k^*_x)^2}.
\end{equation}
With the same accuracy, the extremum energy $\pm uk^{(0)}$ at
$k_x=\lambda k^*_x$ coincides with the extremum energy at $k_x=0$ in
the symmetric quantum well, and a difference between ones $\delta E$
is a higher-order infinitesimal, namely, $|\delta E|\ll
u|k^*_x|\ll\Delta$. Expanding \eqref{15} to the second-order
infinitesimal and remaining terms with $k^{*2}_x$ and $k^*_xV$ as
terms of the same infinitesimal, we obtain
\begin{equation}\label{17}
k^*_x=\frac{V}{\overline{u}}\cdot\frac{\Delta}{F}\cdot\left[1+\sqrt{1+
\frac{u\overline{u}k^2_0}{(\Delta^2-u^2k^2_0)^{3/2}}F}\right],
\end{equation}
where
$F=2\sqrt{\Delta^2-u^2k^2_0}+\frac{u\overline{u}k^2_0}{\sqrt{\Delta^2-u^2k^2_0}}$.

{\bf\textit{The nonsymmetric quantum well of the second type.}}

So we refer to the quantum well with $\Delta_1\neq\Delta_3,\,
V_1=V_3=V,\, V_2=0,\, \Delta_2=0,\, u_1=u_3=\overline{u},\, u_2=u$.
We will suppose that a deviation from the symmetric quantum well is
small $\Delta_1-\Delta_3=2\Delta^\prime,\,
|\Delta^\prime|\ll\Delta_{1,3}$. It is convenient to introduce the
half-sum of the energy gaps
$\widetilde{\Delta}=\frac{\Delta_1+\Delta_3}{2}$, and the energy
gaps are expressed in terms of $\widetilde{\Delta}$ and
$\Delta^\prime$, $\Delta_1=\widetilde{\Delta}+\Delta^\prime,\,
\Delta_3=\widetilde{\Delta}-\Delta^\prime$.

Analogously to the first type of the quantum well, we find the
approximate solution in the form
\begin{equation}\label{18}
\widetilde{E}^{(1)}_\lambda=\pm
u\sqrt{\widetilde{k}^{(0)2}+(k_x-\lambda k^*_x)^2},
\end{equation}
where $\widetilde{k}^{(0)}$ is the root of equation for the
symmetric quantum well \eqref{11}, in which $\widetilde{\Delta}$ is
taken instead of $\Delta$. Further, by analogy with previous case,
one can obtain the formula
\begin{equation}\label{19}
k^*_x=-\frac{\Delta^\prime}{\overline{u}}\cdot\frac{b+\sqrt{b^2+ac}}{a},
\end{equation}
where
$a=2(\widetilde{\Delta}+u\widetilde{k}_0-V)\sqrt{\widetilde{\Delta}^2-(u\widetilde{k}_0-V)^2}+
\overline{u}\widetilde{k}_0(u\widetilde{k}_0-V)\sqrt{\frac{\widetilde{\Delta}+
u\widetilde{k}_0-V}{\widetilde{\Delta}-u\widetilde{k}_0+V}}$,
$b=((\overline{u}-u)\widetilde{k}_0+V)(\widetilde{\Delta}+u\widetilde{k}_0-V)$,
$c=\overline{u}\widetilde{k}_0(u\widetilde{k}_0-V)\frac{2\widetilde{\Delta}^2-(u\widetilde{k}_0-V)^2}
{(\widetilde{\Delta}-u\widetilde{k}_0+V)\sqrt{\widetilde{\Delta}^2-(u\widetilde{k}_0-V)^2}}$;
$\widetilde{k}_0$ is the root of equation \eqref{11} at $k_x=0$ (we
neglect the dependence of $\widetilde{k}^{(0)}$ on $k_x$).

{\bf\textit{The interface states.}}

The interface states (IS) were first predicted by I. E. Tamm in 1932
\cite{Tamm}. This states are localized near the narrow-gap
semiconductor--graphene interfaces\footnote{We note that analogous
states arise in a single heterojunction with an intersection of the
dispersion curves \cite{Andushin2, Kolesnikov1}. They are not a
specificity of the quantum well and can also arise in the quantum
barriers \cite{Ando}.}. The solution in the ranges $z<0$ and $z>d$
is found in the same form as for the size-quantization states and in
the quantum well in the form
\begin{equation*}
\psi_{\lambda1}=B_1e^{-qz}+\widetilde{B}_1e^{qz},\,
\psi_{\lambda3}=B_3e^{-qz}+\widetilde{B}_3e^{qz}.
\end{equation*}
The quantity $q$ is connected with $E_\lambda$ by the relation
\begin{equation}\label{20}
E^2_\lambda=u^2_2k^2_x-u^2_2q^2,
\end{equation}
therefore, the condition of an IS existence is the
inequality\footnote{Such inequality were considered in Ref.
\cite{Blanter} where term ``evanescent modes'' is used instead of
the term ``IS''.}
\begin{equation}\label{21}
|q|\leq|k_x|.
\end{equation}
The equation in $q$ is analogous to \eqref{10} with only difference
that there is $\tanh qd$ instead of $\tan kd$ and $q$ enters into
$A$ instead of $k$.

In order to valid inequality \eqref{21} for all values of $k_x$, it
is necessary that the solution $q$ to this equation vanishes at
$k_x\rightarrow0$. Otherwise, the IS arise at some critical $k_x$ or
they are not exist. It is simply shown that there is no such point
$k_{x0}\neq0$ that $|q(k_{x0})|=|k_{x0}|$ in the case $q(k_{x})\neq
const$. The function graph $|q(k_x)|$ at $k_x\neq0$ lyes either
lower the straight line $q=|k_x|$ (there exist the IS) or higher the
line $q=|k_x|$ (there exist no the IS). The analysis of the solution
of the equation at small $k_x$ is presented in \hyperlink{Tab}{Table 1}.

In particular, the existence of the IS of one parity is possible. A
consequence of this fact is a spin current along the interface. The
mentioned effect can be used in spintronics \cite{Zutic}.

The IS  can exist for charge carriers of one sign. The IS are
energetically more favorable than the size-quantization states. This
fact results in accumulation of the IS by the charge carriers of one
type. Thus at an optical excitation, a charge of one sign will be
localized along the interface against the background of the
uniformly distributed charge of opposite sign.

Recently, it was shown within the electrostatic approach that the
charge is induced at boundaries of the single GNR on the substrate
at applying of an electric field transversely to its plane
\cite{Efetov}. We think that a similar effect is also possible in
the considered heterostructure due to a predominant occupancy of the
IS.

\newpage
\hypertarget{Tab}{}
Table 1. \textit{The results of the analysis of the solutions of the
equation for the IS with $|q|\leq|k_x|$.}
\begin{center}
\begin{tabular}{|l|l|l|l|}
\hline&\text{There are no the IS}&\text{There are the IS}&\text{There are the IS}\\
&&\text{of only one parity}&\text{of both parities}\\
\hline \text{Electrons}&$\left\{
\begin{aligned}p_1>1\\p_3>1\end{aligned}\right.$&$\begin{aligned}\lambda=+1\\ \left\{
\begin{aligned}p_1\geq1\\p_3\leq1\end{aligned}\right.\end{aligned}\hspace{1cm}\begin{aligned}\lambda=-1\\ \left\{
\begin{aligned}p_1\leq1\\p_3\geq1\end{aligned}\right.\end{aligned}$&$\left\{
\begin{aligned}p_1<1\\p_3<1\end{aligned}\right.$\\&&$\hspace{1.6cm}p_1\neq p_3$&\\
\hline \text{Holes}&$\left\{
\begin{aligned}p_1<1\\p_3<1\end{aligned}\right.$&$\begin{aligned}\lambda=+1\\ \left\{
\begin{aligned}p_1\leq1\\p_3\geq1\end{aligned}\right.\end{aligned}\hspace{1cm}\begin{aligned}\lambda=-1\\ \left\{
\begin{aligned}p_1\geq1\\p_3\leq1\end{aligned}\right.\end{aligned}$&$\left\{
\begin{aligned}p_1>1\\p_3>1\end{aligned}\right.$\\&&$\hspace{1.6cm}p_1\neq p_3$&\\ \hline
\end{tabular}

\textit{Notations $p_1=\sqrt{\frac{\Delta_1+V_1}{\Delta_1-V_1}},\,
p_3=\sqrt{\frac{\Delta_3+V_3}{\Delta_3-V_3}}$ are introduced. The
particular case $p_1=p_3=1$ corresponds to a lack of the IS at small
$k_x$ for both electrons and holes (see the text).}
\end{center}

The case $p_1=p_3=1$ ($\Delta_1=\Delta_3$ and $V_1=V_3=0$)
corresponds to the lack of the IS at small $k_x$ for both electrons
and holes. The IS of both parities can exist at $|k_x|>k_{xc}$. In
the case $u_1=u_2=u_3=u$, it is simply obtained from \eqref{12}
\begin{equation}\label{22}
\tanh qd=-\frac{\sqrt{\Delta^2+u^2q^2}}{uq},
\end{equation}
the right-hand part of \eqref{22} is negative and equation
\eqref{22} has not solutions. Therefore in the case $p_1=p_3=1$, an
existence of the IS is possible only at $u\neq\overline{u}$ and
$|k_x|>k_{xc}$.

Now, let us analyse a formation of the IS at a change of the width
$d$ of the quantum well. They can disappear at some critical width
$d_c$ so that there are no the IS at $d\leq d_c$.

Let us determine a criterion of the existence of the IS at the
defined width $d$. The equation in $q$ has the form
\begin{equation}\label{23}
\tanh x=f(x),
\end{equation}
where $x=qd$ and $f(x)$ tends to an asymptotic $f(x)\simeq x$ at
large $x$. For the existence of the solution to equation \eqref{23},
the following two-sided inequality must be valid
\begin{equation}\label{24}
0<f^\prime(0)<1.
\end{equation}
In the general case of the nonsymmetric quantum well, inequality
\eqref{24} can not be solved analytically and we present the results
for the symmetric quantum well.

Let us consider the case of the symmetric quantum well with $V=0$.
This corresponds to $p_1=p_3=1$. The IS arise only at
$u>\overline{u}$, otherwise $f^\prime(0)<0$. One can see that the
qualitative criterion of the IS existence, the intersection of the
dispersion curves for the conduction and valence bands of the
narrow-gap semiconductor with the dispersion straight lines for
graphene \cite{Kolesnikov1}, is realized. It should be expected that
the IS exist for both electrons and holes. As it was noted above,
the IS arise at momenta of the free motion higher some critical
momentum
\begin{equation}\label{25}
|k_x|>k_{xc}=\frac{1}{\sqrt{2}d}\sqrt{\frac{u+\overline{u}}{u-\overline{u}}}
\left\{\sqrt{1+
\frac{4d^2\Delta^2}{(u+\overline{u})^2}}-1\right\}^{1/2}.
\end{equation}
The IS exist if the inequality $|q(k_{xc})|<|k_x|$ is valid. It
provides along with the condition of the lack of the intersection
$|q(k_x)|$ with the straight line $q=|k_x|$ a fulfilment of
inequality \eqref{21} in the whole range $|k_x|>k_{xc}$.

The fact, that the IS begin at some critical momentum of the free
motion, is in accord with the result obtained for
$\Delta_1=\Delta_2=\Delta_3$ and $V_1=V_2=V_3$ in Ref.
\cite{Kolesnikov2}.

For $V\neq0$ and $u=\overline{u}$, the IS arise at $E_\lambda V<0$.
They exist for electrons when $V<0$, i.e. $p_1=p_3<1$, and for holes
when $V>0$, i.e. $p_1=p_3>1$. There are the intersections of the
straight lines of graphene with the dispersion curves either for the
conduction band (there is the IS for electrons) or for the valence
band of the narrow-gap semiconductor (there is the IS for holes).
The critical momentum is
\begin{equation}\label{26}
k_{xc}=\frac{1}{d^2|V|}\left(-u+\sqrt{u^2+(\Delta^2-V^2)d^2}\right).
\end{equation}

{\bf\textit{The optical transitions.}}

In presence an electromagnetic wave ${\bf A}=\text{\bf e}Ae^{i({\bf
qr}-\omega t)}$ where $\text{\bf e}$ is the polarization vector of
the electromagnetic wave, the Dirac equation takes the form
\begin{equation}\label{27}
\left\{u_i\gamma^0\boldsymbol{\gamma}\left(\widehat{\bf
p}-\frac{e}{c}\bf{A}\right)+
\gamma^0\Delta_i+V_i\right\}\Psi=i\frac{\partial}{\partial t}\Psi.
\end{equation}

The operator
$\widehat{O}=-\frac{e}{c}u_i\gamma^0\boldsymbol{\gamma}{\bf A}$ is considered as a perturbation. The matrix element of this
operator is expressed by the matrix element of the rate $v=\text{\bf
e}\cdot{\bf v}$ for the interband transition from the valence to
conduction band $E_\lambda \rightarrow
E^{\prime}_{\lambda^{\prime}}$ \cite{Idlis}
\begin{equation}\label{28}
{\bf
v}=u_i\langle\Psi_{\lambda^\prime}|\gamma^0\boldsymbol{\gamma}|\Psi_\lambda\rangle,
\end{equation}
where $u_i$ implies that $u_1,\, u_2$ and $u_3$ are used for $z<0$,
for $0<z<d$, and $u_3$ for $z>d$, respectively, at calculation
\eqref{28}.

The calculation of the matrix element of the rate by formula
\eqref{28} for the size-quantization gives
\begin{equation*}
v_x=2\lambda\delta_{\lambda\lambda^\prime}A^*_1A_1\left\{i(a^*-a)\frac{u_1}{2k_1}+iu_2
\left[(b^*_+b-b_+b^*+f^*f(b^*b_--bb^*_-))d+(f^*b^*b_++\right.\right.
\end{equation*}
\begin{equation}\label{29}
\left.\left.+f^*b^*_-b)\frac{e^{2ikd}-1}{2ik}-(fb^*_+b+fb^*b_-)\frac{1-e^{-2ikd}}{2ik}\right]+
i(c^*\widetilde{c}-c\widetilde{c}^*)\frac{u_3}{2k_3}e^{-2k_3d}\right\},
\end{equation}
\begin{equation*}
v_z=2\delta_{\lambda\lambda^\prime}A^*_1A_1\left\{(a+a^*)\frac{u_1}{2k_1}+
u_2\left[(b^*b_++bb^*_+-f^*f(b^*_-b+b_-b^*))d+(f^*b^*_-b-\right.\right.
\end{equation*}
\begin{equation}\label{30}
\left.\left.-f^*b^*b_+)\frac{e^{2ikd}-1}{2ik}+(fb^*b_--fb^*_+b)\frac{1-e^{-2ikd}}{2ik}\right]+
(c^*\widetilde{c}+\widetilde{c}^*c)\frac{u_3}{2k_3}e^{-2k_3d}\right\}.
\end{equation}
The coefficient $A_1$ is determined by normalizing condition
\eqref{9}
\begin{equation*}
2A^*_1A_1=\left(\frac{1+a^*a}{2k_1}+b^*b(1+f^*f)(1+\widetilde{b}^*\widetilde{b})d+
\frac{b^*b}{2k}(f^*(\widetilde{b}^2-1)+f((\widetilde{b}^*)^2-1))\sin(2kd)+\right.
\end{equation*}
\begin{equation*}
\left.+\frac{ib^*b}{2k}\left(f^*((\widetilde{b}^*)^2-1)-f((\widetilde{b}^*)^2-1)\right)\sin^2(kd)+
\frac{c^*c+\widetilde{c}^*\widetilde{c}}{2k_3}e^{-2k_3d}\right)^{-1},
\end{equation*}
where
\begin{equation*}
a=\frac{-iu_1k_1+i\lambda u_1k_x}{E_\lambda+\Delta_1-V_1},\,
b_+=\widetilde{b}b,\, b_-=\widetilde{b}^*b,\,
b=\sqrt{\frac{u_1}{u_2}}\frac{1}{1-f},\,
\widetilde{b}=\frac{u_2k+i\lambda u_2k_x}{E_\lambda},
\end{equation*}
\begin{equation*}
\widetilde{c}=\sqrt{\frac{u_1}{u_3}}\frac{e^{ikd}-fe^{-ikd}}{1-f}e^{k_3d},\,
c=\frac{iu_3k_3+i\lambda
u_3k_x}{E_\lambda+\Delta_3-V_3}\widetilde{c},
\end{equation*}
\begin{equation*}
f=1+\frac{2u_2k(E_\lambda+\Delta_1-V_1)}{(-u_2k+i\lambda
u_2k_x)(E_\lambda+\Delta_1-V_1)+E_\lambda(iu_1k_1-i\lambda u_1k_x)}.
\end{equation*}

For the calculation of the IS matrix element of the rate by formula
\eqref{28}, one should replace $k$ with $iq$
\begin{equation*}
v_x=2\lambda\delta_{\lambda\lambda^\prime}A^*_1A_1\left\{i(a^*-a)\frac{u_1}{2k_1}+iu_2
\left[-bb_+\frac{1-e^{-2qd}}{q}+2fb(b_+-b_-)d+f^2bb_-\frac{e^{2qd}-1}{q}\right]\right.+
\end{equation*}
\begin{equation}\label{31}
\left.+iu_3\widetilde{c}(c^*-c)\frac{1-e^{-2k_3d}}{2k_3}\right\},
\end{equation}
\begin{equation}\label{32}
v_z\equiv0.
\end{equation}
Here, normalizing \eqref{9} results in the expression
\begin{equation*}
2A^*_1A_1=\left(\frac{1+a^*a}{2k_1}+2f(b^*_+b_--b^2)d+(b^2+b^*_+b_+)\frac{1-e^{-2qd}}{2q}+
f^2(b^2+b^*_-b_-)\frac{e^{2qd}-1}{2q}+\right.
\end{equation*}
\begin{equation*}
\left.+(\widetilde{c}^2+c^*c)\frac{e^{-2k_3d}}{2k_3}\right)^{-1}.
\end{equation*}
It is evident from formulae \eqref{29}-\eqref{31} that only the
transitions with the conservation of the parity is allowed
(\hyperlink{fig2}{Fig. 2} shows them by arrows). The optical transitions for the IS
exist only in the particular case $p_1=p_3=1$ at an absorption of
the electromagnetic waves with the linear polarization along the
interfaces.

These conclusions can be experimentally verified by the following
way. Let us take the symmetric quantum well. Applying a voltage $U$
to the GNR, one can achieve that the work function $V$ becomes equal
to zero, i.e. the particular case $p_1=p_3=1$ is realized at
$U\neq0$. Then the optical transitions arise. They are due to the
absorption of the electromagnetic waves with an energy $E$ so that
$E_{min}<E<\Delta_{eff}$ where $\Delta_{eff}$ is the effective
energy gap in the size-quantization spectrum (the smallest energy
difference between the electron and hole branches).

\newpage

{\bf\textit{The possibility of the exciton formation.}}

In conclusion, we note that a formation of the finite effective
energy gap in the size-quantization spectrum can result in the
exciton generation at the optical pumping (the energy gap in graphene
is zero and there are no excitons).

The ortoexcitons with the total spin $S=1$ of an electron and a hole
are direct in the the symmetric quantum well therefore they are
short-lived, and the paraexcitons with $S=0$ \cite{Silin2} are
long-lived, since the transitions with the change in the parity are
forbidden (out of the dependence wether they are direct or not). The
ortoexcitons are indirect in the nonsymmetric quantum well
(\hyperlink{fig2}{Fig. 2}), therefore they must also be long-lived.
It was noted earlier that the paraexcitons are short-lived for the
layered narrow-gap semiconductor heterostructures because the
transitions with the change of the parity are allowed
\cite{Ratnikov2}.

The calculation of exciton binding energy in the planar quantum well
based on graphene and the narrow-gap semiconductors represents a
separate problem and it will be present elsewhere.

\vspace{2cm}
\begin{center}
\hypertarget{fig2}{}
\includegraphics[width=10cm]{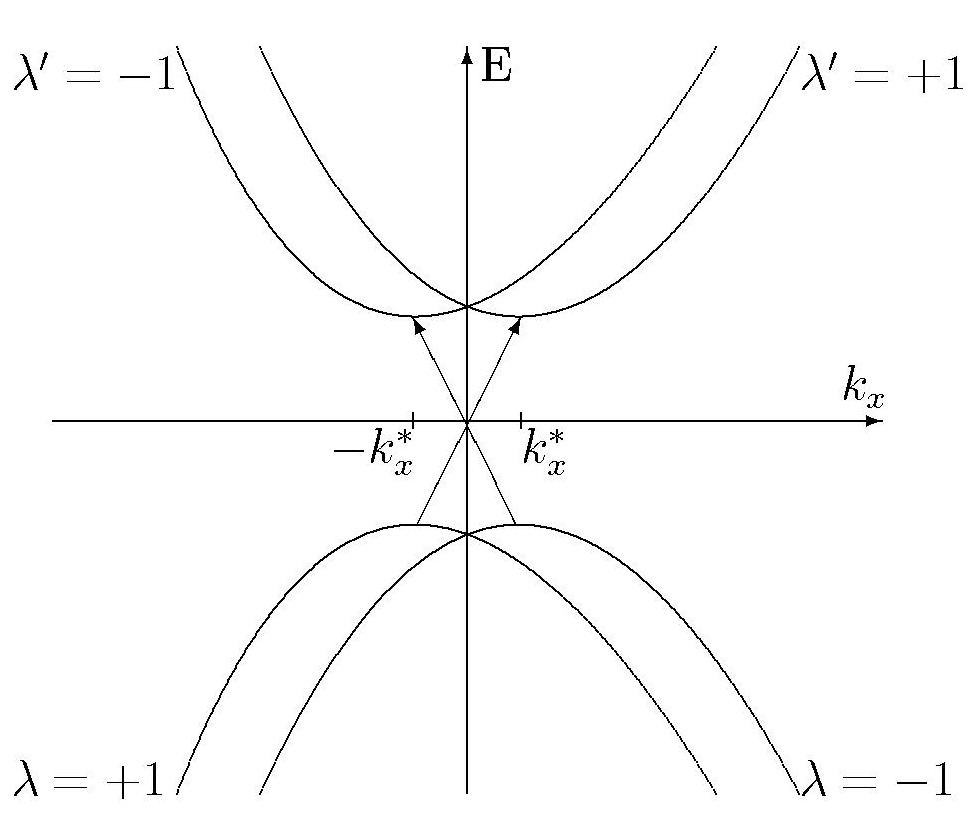}

\vspace{1cm}
Fig. 2. \textit{Splitting of the energy spectrum in the nonsymmetric
quantum well. The allowed transitions with the conservation of the
parity are shown.}
\end{center}

\end{document}